\documentclass{ws-ijmpa}
\def\halftext{.450\textwidth}
\newcommand{\kp}{K^+}

\newcommand{\kn}{K^-}

\newcommand{\g}{\gamma}

\newcommand{\etap}{\eta^{\prime}}

\newcommand{\jpsi}{J/\psi}
\newcommand{\psip}{\psi^{\prime}}

\newcommand{\pip}{\pi^+}

\newcommand{\pin}{\pi^-}
\newcommand{\pio}{\pi^0}

\newcommand{\gff}{\gamma f_{2} f_{2}}

\newcommand{\etac}{\eta_{c}}
\newcommand{\ar}{\rightarrow}

\newcommand{\bfg}{\begin{figure}[htpb]}
\newcommand{\efg}{\end{figure}}
\newcommand{\bitm}{\begin{itemize}}
\newcommand{\eitm}{\end{itemize}}
\newcommand{\bnum}{\begin{enumerate}}
\newcommand{\enum}{\end{enumerate}}
\newcommand{\btbl}{\begin{table}[htp]}
\newcommand{\etbl}{\end{table}}
\newcommand{\btbu}{\begin{tabular}[htp]}
\newcommand{\etbu}{\end{tabular}}

\newcommand{\bcl}{\begin{center}}
\newcommand{\ecl}{\end{center}}

\newcommand{\beq}{\begin{equation}}
\newcommand{\eeq}{\end{equation}}
\newcommand{\beqr}{\begin{eqnarray}}
\newcommand{\eeqr}{\end{eqnarray}}
\begin{document}

\markboth{Shuangshi Fang} {Recent Results on $\jpsi$ Decays from BESII}

\catchline{}{}{}{}{}
%
%%%%%%%%%%%%%%%%%%%%%%%%%%%%%%%%%%%%%%%%%%%%%%%%%%%%%%%%%%%%%%%%%%%%

\title{Recent Results on $\jpsi$ Decays from BESII}

\author{\footnotesize Shuangshi Fang\\
(For the BES Collaboration)
%\footnote{
%Typeset names in 8 pt roman, uppercase. Use the footnote to indicate the
%present or permanent address of the author.}
}

\address{China Center for Advanced Science and Technology(CCAST),\\
Beijing 100080, People's Republic of China\\
E-mail:~fangss@mail.ihep.ac.cn\\
The International Conference on QCD and Hadronic Physics\\
16-20~Jun., 2005, Peking University, Beijing, China
%\footnote{State completely without abbreviations, the
%affiliation and mailing address, including country. Typeset in 8 pt
%italic.}
 }

%\author{SECOND AUTHOR}
%
%\address{Group, Laboratory, Address\\
%City, State ZIP/Zone, Country
%}

\maketitle

%pub{Received (Day Month Year)}{Revised (Day Month Year)}

\begin{abstract}
The new observation of X(1835)is  reported using 58 million $\jpsi$ events
collected at BES II detector.
We also present the measurements of $\jpsi$ and $\eta_c$ decays, of them
 some
are the first measurements and some improve the precision
in previous measurements.

\end{abstract}

\section{INTRODUCTION}
Since the discovery of the
$\jpsi$ at Brookhaven~\cite{aube} and SLAC~\cite{augu} in 1974, more
than one hundred exclusive decay modes of the $\jpsi$ have been
reported. The $\jpsi$ decays provide an excellent source of events 
to study light hadron spectroscopy and search for glueballs, hybrids,
and exotic states.
Recently, $5.8\times 10^{7}$ $\jpsi$ events 
have been obtained with the upgraded Beijing
Spectrometer (BESII)\cite{BESII}. 
Many important results on the search for multiquark, 
study oc the light scalar mesons and 
excited baryon states,
as well as the measurements of $\jpsi$ decays are performed based on this
data sample.

\section{New observation of X(1835)}

\begin{figure}[htb]
\parbox{\halftext}{\psfig{file=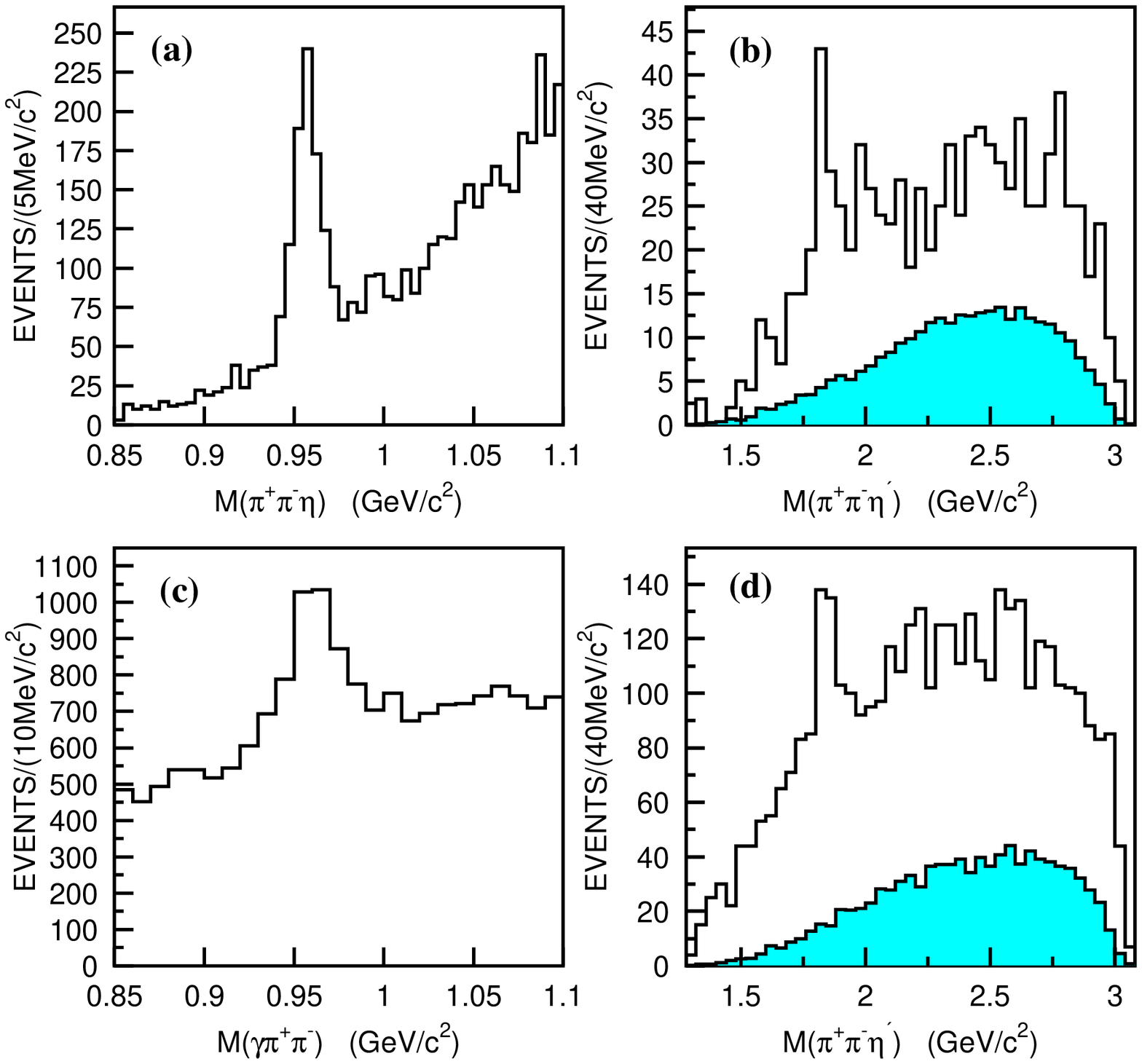,width=\halftext,height=4.5cm}
\caption{
Invariant mass distributions for selected
           $\jpsi\ar\g\pip\pin\etap$ candidate events:
           (a) The $\pip\pin\eta$ invariant mass distribution.
           (b) The $\pip\pin\etap$ invariant mass distributions.
           (c) The $\g\pip\pin$ invariant mass distribution.
           (d) The $\pip\pin\etap$ invariant mass distributions.
               The open histograms are data and the shaded histograms
               represent
               $\jpsi\ar\g\pip\pin\etap$
               phase-space MC events (with arbitrary
normalization).}\label{x1835dis}}
         \hspace{2mm}
\parbox{\halftext}{\psfig{file=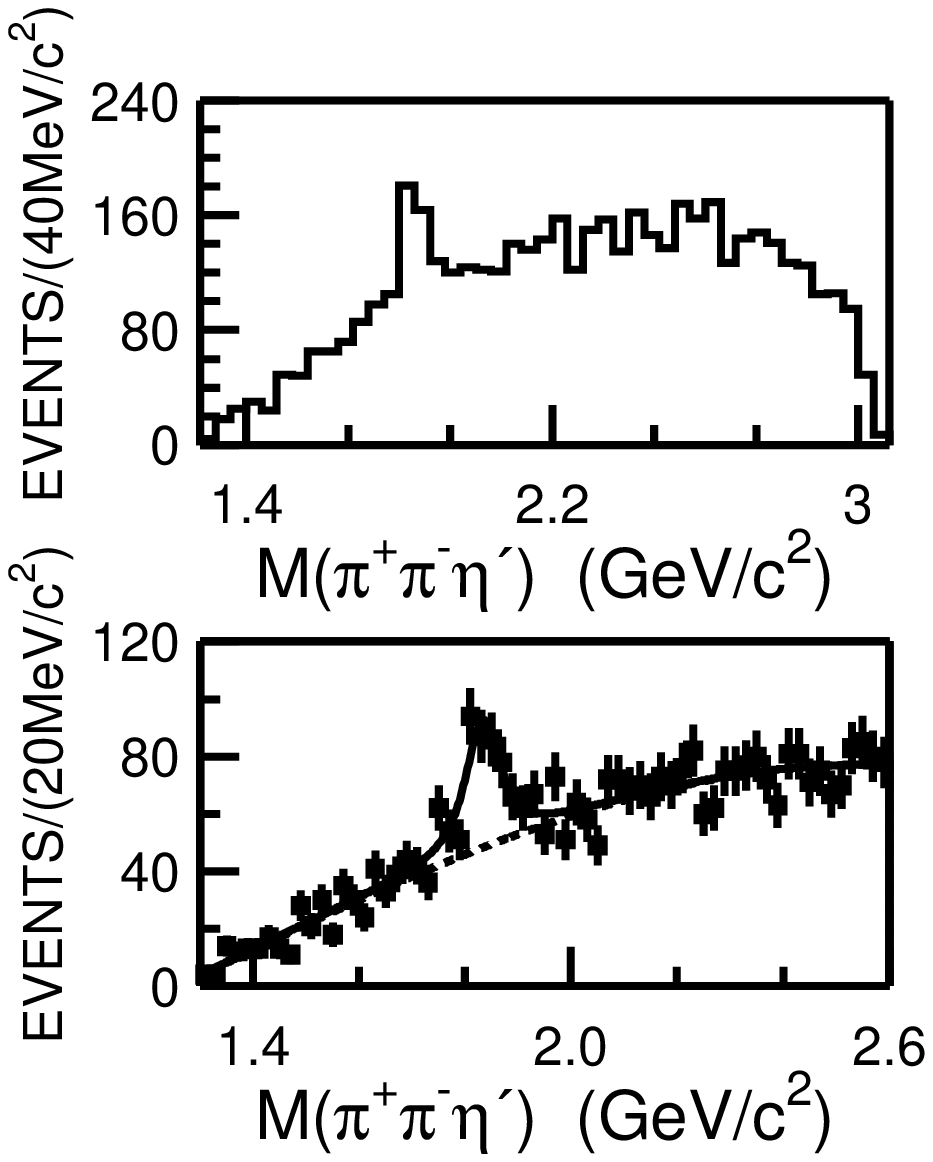,width=\halftext,height=6.0cm}
\caption{ The $\pip\pin\etap$ invariant mass distribution for
           selected events from the
           $\jpsi\ar\g\pip\pin\etap(\etap\ar\pip\pin\eta,\eta\ar\g\g)$
           and $\jpsi\ar\g\pip\pin\etap(\etap\ar\g\rho)$ analyses.
           The bottom panel shows the fit (solid curve) to the data (error
           bars); the dashed curve indicates the background
function.}\label{x1835fit}}
\end{figure}

In this letter, we report the first
analysis on the $\jpsi\ar\g\pip\pin\etap$ decay channel. In the analysis,
$\etap$ is tagged in its two decay modes,
$\etap\ar\pip\pin\eta (\eta\ar\g\g)$ and $\etap\ar\g\rho$.
igure~\ref{x1835dis}(a) shows the $\pip\pin\eta$ invariant mass distribution
 and the $\etap$ signal is seen. The  $\pip\pin\etap$ invariant mass
spectrum for the selected events is shown in Fig.~\ref{x1835dis}(b), where
a peak at a mass around 1835 MeV/c$^2$ is observed.
The $\g\pip\pin$ invariant mass distribution shows a clear
$\etap$ signal (Fig.~\ref{x1835dis} (c)).
A  peak near 1835~MeV/c$^2$ is also evident in the $\pip\pin\etap$
invariant mass spectrum (Fig.~\ref{x1835dis} (d)).

Figure~\ref{x1835fit} shows the $\pip\pin\etap$ invariant mass spectrum for the combined 
$\jpsi\ar\g\pip\pin\etap$ $(\etap\ar\pip\pin\eta)$ and
$\jpsi\ar\g\pip\pin\etap$ $(\etap\ar\g\rho)$ samples.
A clear peak near 1835~MeV/c$^2$ is observed. To ensure that the peak near
1835~MeV/c$^2$ is not due to background,
we have made extensive studies of potential background processes
using both data and MC.  Non-$\etap$ processes are studied
with $\etap$ mass-sideband events.   The main background channel,
$\jpsi\ar\pio\pip\pin\etap$,  and other background processes with
multi-photons and/or with kaons are reconstructed with the data.
In addition, we also checked for possible backgrounds with a
MC sample of 60 million $J/\psi$ decays generated according
to the LUND model~\cite{chenjc}.   None of these backgrounds
are found to produce a peak around 1835 MeV/c$^2$ in the
$\pip\pin\etap$ invariant mass spectrum.

This spectrum is fitted with a
Breit-Wigner (BW) function convolved with a Gaussian mass resolution
function (with $\sigma = 13$~MeV/c$^2$) to represent the
$X(1835)$ signal plus a smooth polynomial
background function.
The BW mass and width obtained from the fit (shown in the bottom panel of
Fig.~\ref{x1835fit}) are  $M=1833.7\pm 6.1\pm2.7$ MeV/c$^2$ and 
$\Gamma=67.7\pm 20.3\pm7.7$
MeV/c$^2$. The BW signal yield from the fit is $264\pm 54$ events and
the statistical significance for the signal is 7.7~$\sigma$.
The mass and width of the $X(1835)$ are
not consistent with any known particle~\cite{pdg}.
 While the measured
$X(1835)$ mass is consistent with the
mass obtained from the $\jpsi\ar\g p\bar{p}$ channel, the
measured width is higher by $1.9\sigma$ than the upper limit
on the width reported in Ref.~\cite{gpp}.
Thus, the $X(1835)$ is a candidate
for the resonance that produces the $p\bar{p}$ mass threshold
enhancement in $\jpsi\ar \g p\bar{p}$ process.

Using MC-determined selection efficiencies
of $(3.72\pm 0.06)\%$ and $(4.85\pm0.07)\%$
for $\etap\ar\pip\pin\eta$ and $\etap\ar\g\rho$ modes, respectively,
we find the product branching fraction to be:

\begin{center}
$B(\jpsi\ar\g X(1835))\cdot B(X(1835)\ar\pip\pin\etap)= (2.2\pm 0.4\pm 0.4)\times
10^{-4}$
\end{center}
\section{EXCITED BARYON STATES}\label{subsec:fig}
\begin{figure}[htbp]
\centerline{ \hbox{\psfig{file=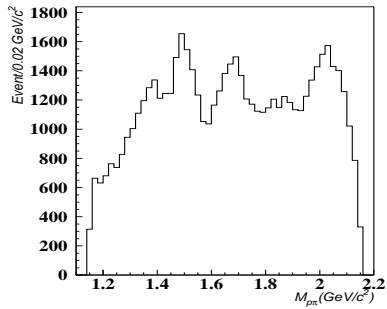,width=5.0cm,height=4.0cm}}}
\caption{The $p\pi$ invariant mass distribution for $\jpsi \ar
{p}\pi^{-}\bar{n}$.
}
\label{nstar}
\end{figure}

The analysis on a study of $N^*$ resonances was reported in
Ref.~\cite{pnpi}. More than 100 thousand
$\jpsi\ar p\pi^{-}\bar{n}+c.c.$ candidate events are obtained and the
backgound level is about 8\%. There are four clear peaks around 1360
MeV/$c^2$, 1500 MeV/$c^2$,
1670 MeV/$c^2$ and 2065 MeV/$c^2$ observed in  the $p\pi$ invariant mass
spectrum, as shown in Fig. \ref{nstar}.
They are the first direct observation of the $N^{*}(1440)$ peak
and a long-sought "missing" $N^{*}$ peaks above 2 GeV in the $\pi
N$ invariant mass spectrum. A simple Breit-Wigner fit gives the
mass and width for the $N^{*}(1440)$ peak as $1358\pm6\pm16$  MeV/$c^2$
and $179\pm26\pm50$ MeV/$c^2$, and for the new $N^{*}$ peaks above 2 GeV
as $2068\pm3^{+15}_{-40}$ MeV/$c^2$ and $165\pm14\pm40$ MeV/$c^2$.

\section{MEASUREMENTS OF SOME $\jpsi$ DECAYS}
The largest $\jpsi$ sample, 58 million $\jpsi$ events, offers 
a unique opportunity to
measure precisely the branching fractions of $\jpsi$ decays.
The largest $J/\psi$ decay involving hadronic resonances is $J/\psi
 \ar \rho(770) \pi$.  Its branching fraction is measured to be
$(21.84\pm0.05\pm2.01)\times 10^{-3}$\cite{rhopi} which is obviously higher than
the PDG value,$(1.5\pm0.2)\%$. To check this result, we present another
independent measurements of this branching fraction  obtained from a comparison
of the rates for $\psip\ar\pip\pin\jpsi$, $\jpsi\ar\pip\pin\pio$ and
$\jpsi\ar\mu^+\mu^-$.
The result is  $(20.91\pm0.21\pm1.16)\times 10^{-3}$ which is consistent
with that obtained from direct $\jpsi$ decays. The weight mean of these two
measurements is $(2.10\pm0.21)\%$ which is in good agreement with the
subsequent result, $(21.8\pm1.9)\%$\cite{babar}, from BaBar Collaboration.
Then we presented the branching fractions of
$\jpsi\ar\phi\pio,\phi\eta,\phi\etap$ which are shown in
Table.~\ref{brpv}\cite{brphip}.
\begin{table}[htpb]
\caption{Branching fractions of $\jpsi\ar\phi\pio$, $\phi\eta$,
 and $\phi\etap$.}
\begin{center}
\begin{tabular}{ l |c|c }
\hline
\hline
$\jpsi\ar$          & Final states &B.F.
($\times 10^{-4}$) \\
\hline
$\phi\pio$& $\kp\kn\g\g$ & $<0.064$ (C.L. 90\%)
%{To conservatively estimate the upper limit, the result obtained from
%formula (1) is corrected by dividing a factor $(1- \sigma_{sys})$. Here,
%$\sigma_{sys}$ is the systematic error for this decay mode.}
 \\

\hline
             &$\kp\kn\g\g$ & 8.67$\pm$0.19$\pm$0.93 \\
             & $\kp\kn\pip\pin\g$ & 9.79$\pm$1.02$\pm$1.17\\
$\phi\eta$    & $\kp\kn\pip\pin\g\g$ & 9.41$\pm$0.30$\pm$1.19 \\
              & Average   &8.99$\pm$0.18$\pm$0.89 \\
              & PDG       & $6.5\pm0.7$ \\
\hline
              &$\kp\kn\g\g$ & 6.10$\pm$1.34$\pm$0.73\\

$\phi\etap$  & $\kp\kn\pip\pin\g$ & 5.64$\pm$0.35$\pm$0.70 \\

              & $\kp\kn\pip\pin\g\g$ & 5.11$\pm$0.31$\pm$0.65\\
              & Average              & 5.40$\pm$0.25$\pm$0.56\\
              & PDG                  & $3.3\pm 0.4$ \\
\hline
\hline
\end{tabular}
\end{center}
\label{brpv}
\end{table}

In this talk, we also reported the search the new hadronic decay modes of 
$\jpsi$ decays. $\jpsi\ar\gff$, $\jpsi\ar 2(\pip\pin)$ and $\jpsi\ar
3(\pip\pin)\eta$ are observed and their branching fractions are 
measured for the first time to be
$B(\jpsi\ar\gff)=(9.5\pm0.7\pm1.6)\times10^{-4}$, 
$B(\jpsi\ar 2(\pip\pin)\eta)=(2.26\pm0.08\pm0.27)\times 10^{-3}$, and 
$B(\jpsi\ar 3(\pip\pin)\eta)= (7.24\pm0.96\pm1.11)\times 10^{-4}$.

The large $\jpsi$ sample also provides a chance to observed new decay modes
of $\eta_c$ through $\jpsi\ar\g\etac$. In the analysis, the decays of
$\eta_c$ to $\kp\kn 2(\pip\pin)$ and $3(\pip\pin)$ are
observed for the first time. The product branching fractions are determined
to be
$B(\jpsi\ar\g\etac)\cdot B(\etac\ar\kp\kn\pip\pin\pip\pin)$ $=
(1.21 \pm 0.32\pm
0.23)\times 10^{-4}$ and $B(\jpsi\ar\g\etac)\cdot
B(\etac\ar\pip\pin\pip\pin\pip\pin)=
(2.59 \pm 0.32\pm0.48)\times 10^{-4}$, respectively. The upper limits
of $\etac\ar\phi
\pip\pin\pip\pin$   is also obtained as
$B(\jpsi\ar\g\etac)\cdot
B(\etac\ar\phi\pip\pin\pip\pin)< 6.03\times 10^{-5} $ at 90\% confidence
level.
\section{SUMMARY}
Based on $5.8\times 10^7$ $\jpsi$ events accumulated at the BESII detector,
a resonance around 1835 MeV/$c^2$ is observed in $\jpsi\ar\g\pip\pin\etap$ 
which could be the 
same struncture observed in $\jpsi\ar\g p\bar{p}$. We also report the
measurements of $\jpsi$ and $\eta_c$ decays, of them some
are the first measurements and some improve the precision
in previous measurements.

%\section{Acknowledgments}
%We acknowledges the staff of BEPC for their hard efforts.
%This work is supported in part by the National Natural Science Foundation
%of China under contracts Nos. 19991480, 10225524, 10225525, 10425523, the
%Chinese Academy of Sciences under contract No. KJ 95T-03, the 100 Talents
%Program of CAS under Contract Nos. U-11, U-24, U-25, and the Knowledge
%Innovation Project of CAS under Contract Nos. KJCX2-SW-N10, U-602, U-34
%(IHEP); by the National Natural Science Foundation of China under Contract
%No. 10175060 (USTC); and by the Department of Energy under Contract No.
%DE-FG03-94ER40833 (U Hawaii).

\end{document}